# UAV Networks Surveillance Implementing an Effective Load-Aware Multipath Routing Protocol (ELAMRP)

**Raja Vavekanand** [*], **Kira Sam** [*], **Vijay Singh** [†]
[*]Datalink Research and Technology Lab
[†]Benazir Bhutto Shaheed University Lyari

*Abstract*— In this work uses innovative multi-channel load-sensing techniques to deploy unmanned aerial vehicles (UAVs) for surveillance. The research aims to improve the quality of data transmission methods and improve the efficiency and reliability of surveillance systems by exploiting the mobility and adaptability of UAVs does the proposed protocol intelligently distribute network traffic across multiple channels, considering the load of each channel, While addressing challenges such as load balancing, this study investigates the effectiveness of the protocol by simulations or practical tests on The expected results have improved UAV-based surveillance systems, more flexible and efficient networks for applications such as security, emergency response and the environment alignment of monitoring -Offering infrastructures, which contribute to efficient and reliable monitoring solutions.

## I. INTRODUCTION

Unmanned Aerial Motors (UAVs), also called drones, are remotely piloted planes that can carry out numerous responsibilities without human intervention. UAVs have been broadly used for surveillance programs in numerous domains, along with army, disaster management, and environmental monitoring. Surveillance is looking at and gathering statistics approximately a target or hobby, along with an enemy, a disaster area, or a natural world habitat. UAVs can provide a decision, actual-time, and continual surveillance capability that could decorate the situational cognizance and choice-making of the customers.

However, the communication challenges of UAV networks, such as limited bandwidth, dynamic topology, and high mobility, pose significant problems in ensuring reliable data transmission well within Limited bandwidth means UAVs must compete for scarce wireless resources to transmit and receive data. The dynamic topology means that UAVs must constantly update their routing information to cope with frequent changes in network connectivity. High mobility means that UAVs have to cope with large variations in link quality and interference from other nodes. These challenges can lead to increased network load, long end-to-end latency, low packet delivery ratio, and low throughput, which can degrade the performance and quality of the monitoring service.

*Corresponding Author: Raja Vavekanand*
*Email: bharwanivk@outlook.com*
*Code: GitHub*

Therefore, it's far crucial to layout a powerful routing protocol for UAV network surveillance, which can stability the community load, reduce the end-to-cease put-off, and enhance the packet delivery ratio. A routing protocol is a set of policies and algorithms that decide how the UAVs ahead the facts packet to their locations.

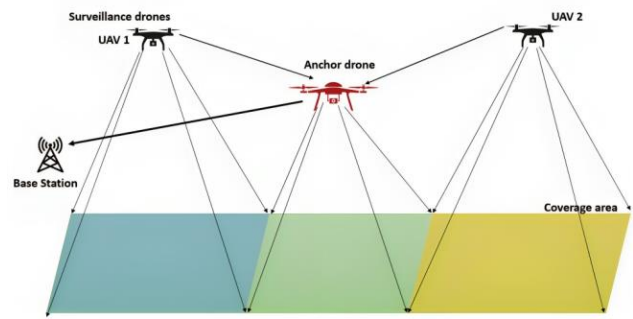

Fig. 1. UAV Surveillance System

A routing protocol can be classified into two categories: single-route and multipath. An unmarried-route routing protocol selects only one course for each records waft, while a multipath routing protocol selects a couple of paths for every information float. A multipath routing protocol can provide several benefits over a single-path routing protocol, together with load balancing, fault tolerance, and bandwidth aggregation. In this record, we present an powerful load-conscious multipath routing protocol (ELAMR) for UAV community surveillance, which aims to stability the network load, lessen the cease-to-give up delay, and improve the packet delivery ratio. ELAMR makes use of a unique load estimation metric that considers both the queue duration and the variety of packets in transmission at every UAV node. ELAMR additionally employs a multipath creation set of rules that selects the paths with the minimum load and the maximum hyperlink fine. We compare the performance of ELAMR through large simulations and compare it with existing routing protocols. The results show that ELAMR outperforms the prevailing protocols in phrases of network load, end-to-stop postpone, packet shipping ratio, and throughput. ELAMR can be implemented for numerous UAV network surveillance eventualities and offers a reliable and efficient communication solution.

The main contributions of this paper are as follows:

1) The approach of strategically placing drones to cover the given area with zero blind spots ensures efficient data transmission.
2) Signal-to-noise ratio and channel capacity calculations help in assessing the quality of communication.
3) Our proposed project helps to monitor battery life at each stage of data transmission.
4) Our proposed methodology ensures reliable and efficient data transmission in a drone network, making it suitable for various applications.

## II. RELATED WORK

Wireless sensor networks (WSNs) face the project of power green records transmission due to the confined battery existence of the sensor nodes. One way to conquer this mission is to use relay nodes that can forward the aggregated records from the sensor nodes to a valuable base station (BS). However, the relay nodes might also have a limited transmission variety and can be unable to establish dependable links with the BS or different relays [1]. Moreover, the relay nodes may additionally devour more strength than the sensor nodes, decreasing the network lifetime.

To address these issues, some researchers have proposed to apply a cellular device-to-device (D2D) communique as a relay mechanism for WSNs. D2D conversation allows the devices to talk at once with every other using mobile resources, without going through the BS. This can lessen the transmission distance, decorate the hyperlink fine, and improve the spectral performance [2] . Furthermore, D2D verbal exchange can offer dual connectivity, which means that the devices can use both mobile and D2D links simultaneously, which could increase the reliability and robustness of the community [1].

The research under review describes a cellular D2D-assisted relay communication strategy for wireless sensor networks (WSNs), in which data is forwarded to the base station (BS) by the cluster heads (CHs) of sensor nodes via D2D relays [3]. The Maximum Weight Bipartite Matching method is used by the proposed strategy, DSPA (D2D Relay Node (DRN) Selection and uplink Power Allocation), to choose the best DRNs for each CH based on the channel gain and the residual energy of the DRNs [4]. Next, taking into account the interference from other pairings as well as the DRNs' energy usage, the strategy allows the uplink power for every CH-DRN pair. In comparison to alternative schemes, the authors demonstrate that their suggested method can increase the energy efficiency, hop gain, and network longevity of the WSNs [5].

Other related works on D2D relay verbal exchange for WSNs encompass [Social Reliable D2D Relay for Trustworthy Paradigm in 5G Wireless Networks], wherein the authors use social relationships amongst devices to shape private accept as true with and to select dependable relays; [Energy-Efficient Power Allocation and Relay Selection Schemes for Relay Assisted D2D Communications in 5G Networks], where the authors suggest schemes for electricity allocation and relay choice, based totally at the sign-to-noise ratio (SNR) and the electricity harvesting functionality of the devices; and [Relay Selection-Based Energy Efficiency of Hybrid Device To-Device Enabled 5G Networks], wherein the authors introduce a hybrid relay community version that uses reactive and proactive relay choice strategies to enhance the electricity efficiency of the network [6].

This review summarizes the main conclusions of a recent study on power regulation and cooperative UAV channel modeling for 5G IoT networks. In a 5G IoT scenario where numerous UAVs work together with cellular base stations, this study investigates the communication problem [7]. It focuses on uplink channel modeling and the investigation of uplink transmission performance. The effects of multi-UAV reflection and 3D distance on wireless signal propagation are considered in the channel model [8]. The three-dimensional distance is used to determine the path loss, which is a more realistic representation of the actual path loss. The UAV's uplink transmit power is modified using the power control factor to account for differences in propagation route losses in order to achieve precise power management [1].

One of the tasks we have in mind is figuring out the optimal UAV grid topology with the shortest drone range for video monitoring over an unreachable area without cellular coverage. This study acquired zero blind spots during surveillance by utilising the concept of the subject of view (FOV) of the onboard camera [4]. Additionally, a wireless community for the UAV grid was established. Each UAV had a Raspberry Pi installed, and the inter-UAV distance was fixed primarily on the basis of excellent quality of experience (QoE) and ideal packet loss for streaming videos. It verified that the video streaming QoE is lower than allowable thresholds at 4

The goal of the work on an energy-efficient UAV surveillance scheme that makes use of the compressive sensing (CS) concept is to minimize the overall energy consumption of the UAV network while meeting the surveillance quality requirement [10]. This is achieved by formulating the UAV surveillance problem as a joint optimization problem of UAV trajectory design and CS measurement matrix design. Through simulations, it proved the efficacy of their plan and used the alternate direction method of multipliers (ADMM) algorithm to solve the optimization problem.

## III. PROPOSED METHODOLOGY

After carefully examining many active projects and previously proposed frameworks (as mentioned in Section II) for efficient UAV network surveillance, we have created a system to improve the current methods used in this field. UAV network surveillance using effective load-aware multipath routing protocol is our proposed methodology.

*A. Algorithm for Effective UAV-based Network Surveillance*

In this scenario, the objective is to optimize the communication between a set of drones within a specified area. The process begins by defining the coverage area of the drones and obtaining the latitude and longitude coordinates from a map. These coordinates are then translated into a coordinate system for further analysis. Next, the task involves strategically placing drones to eliminate blind spots, ensuring comprehensive coverage. The anchor drone is positioned at the centroid, facilitating effective communication with other drones in the network. To enhance communication reliability, the log distance path loss is calculated using the given formula. This computation involves factors such as transmitted power, transmitter, and receiver gains, wavelength, and the path loss exponent for free space.

- **Log Distance Path-Loss:**

$$PL_{LD}(d) = PL_{LD}(d_0) + 10n\log(d/d_0)$$
$$PL_{LD}(d_0) = 10\log[(P_t(G_t G_r^2))/(4d_0)^2] + 10n\log(d) \quad (1)$$

where,
$P_t$ = Transmitted power = $10^{-0.4}$ W
$G_t$ = Transmitted gain = 10
$G_r$ = Receiver gain = 10
$\lambda = c/f$
n = Path loss exponent for free space

Following this, the uplink path loss is determined, considering the time delay and path loss gain. Subsequently, the Signal-to-noise Ratio (SNR) is calculated, incorporating the transmitted power, path loss gain, and the power of additive white Gaussian noise at the receiver.

- **Uplink Path Loss:**

$$\zeta = \tau . d^\alpha \quad (2)$$

where, $\alpha$ = Path loss exponent $\tau$ = Channel coefficient of wireless channel between the $T_X$ and $R_X$

$$R_x = {}^p\beta(d) \quad (3)$$

$\beta(d)$ = Large scale path loss

$$\beta(d) = -PL_{LD}(d)[dB]$$
$$SNR = Pt.\zeta^{-1}.g/N \quad (4)$$

where, g = Rayleigh
fading, $\zeta^{-1}$ = Path loss
gain,
N = Power of AWGN at the receiver = 1.4332 x $10^{-15}$ W

The channel capacity, representing the maximum data rate that can be reliably transmitted, is then computed using the given bandwidth. This capacity is essential for determining the efficiency of data transmission within the network.

- **Channel Capacity:**

$$C = B.\log(1+SNR) \quad (5)$$

where,
B = Bandwidth of the line = 360KHz

To optimize energy consumption, only a percentage of drones (e.g., 10%, 20%, or 30%) are activated at a time to transmit data to the anchor drone. This staggered activation helps in balancing the load and maximizing the lifespan of the drone batteries. Multiple possible communication paths are evaluated, and the most optimum path is selected based on the calculated parameters.

Along with the channel capacity, the battery life of each UAV is calculated after each transmission. For this calculation, the initial battery life of each UAV is assumed between 80% to 100%. The drone configuration to be undertaken for the project is 11.1V and 2200mAH battery power. The actual channel capacity including battery is determined as below:

$$C_{actual} = C_{max}.(Battery_{remaining}/Battery_{max})$$
$$Time_{transmission} = Data_{transmission}/C_{remaining}$$
$$Battery_{max} = (11.1 \times 2200 \times 3.6)J \times Battery_{initial}$$
$$Battery_{remaining} = Battery_{prev} - 10^{-0.4} \times Time_{transmission} \quad (6)$$

Finally, according to the above calculations, the data is transmitted using the optimal path. Then battery life and channel capacity is updated accordingly after each transmission. This approach ensures efficient and reliable communication within the UAV network.

*B. Algorithms*

---

**Algorithm 1** Function for All path

---

1: function FIND_ALL_PATHS_TH(drone graph, start, end, threshold, path)
2:     *path.append*(*start*)     ▷ Add start to path
3:     if *start* = *end* then
4:         return *path*
5:     end if
6:     if *start* not in *drone graph* then
7:         return []
8:     end if
9:     *paths* ← []
10:     for *node* in *drone graph*[*start*] do
11: if *node* not in *path* and len(*path*) < threshold then
12:         *paths*+ = find all paths th(
13:         *drone graph, node, end, threshold,* path.copy()
14:         )
15:     end if
16:     end for
17:     return *paths*
18: end function

```
Algorithm 2 Main Algorithm
 1: trans_cap_of_first_drone ← 6000000
 2: for_one ← find_all_paths_th(drone_graph, 4, 24, 5, [])
 3: cap ← {}
 4: for i in range(len(for_one)) do
 5:     row_wise ← {}
 6:     for j in range(len(for_one[i]) − 1) do
 7:         row_wise[for_one[i][j]] ←
 8:         drone_graph[for_one[i][j]][for_one[i][j + 1]]
 9:     end for
10:     cap[i] ← row_wise
11: end for
12: path_one ← []
13: remaining_data ← 6000000
14: while remaining_data ≥ 0 do
15:     sorted_min_cap ← {}
16:     for key, value in cap do
17:         sorted_min_cap[key] ← min(value.values())
18:     end for
19:     sorted_min_cap ←sorted(sorted_min_cap.items(),
    key=lambda item: item[1], reverse=True)
20:     key, value ← next(iter(sorted_min_cap))
21:     remaining_data ← remaining_data − value
22:     path_one.append(for_one[key])
23:     for k, min_capacity in cap[key].items() do
24:         cap[key][k] ← cap[key][k] − min_capacity
25:         drone_graph[key][k] ←
26:         drone_graph[key][k] − min_capacity
27:     end for
28: end while
```

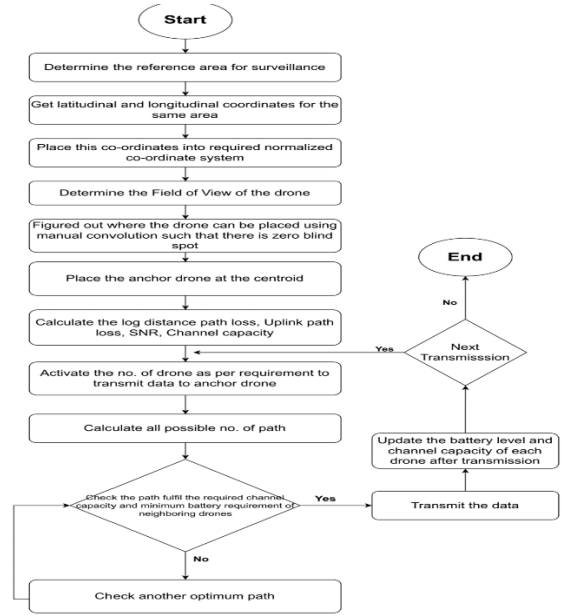

Fig. 2. Flow chart of UAV-based network surveillance

## VI. IMPLEMENTATION AND SIMULATION OF UAV NETWORK

Designing and implementing a surveillance system involves a methodical process, starting with identifying the surveillance area and obtaining geographic coordinates. Subsequently, determining the drone's field of view ensures optimal positioning to eliminate blind spots. Activation of the anchor drone initiates communication coordination among all drones. Calculating transmission parameters like log distance path loss and signal-to-noise ratio aids in selecting the best data transmission paths. Activating necessary drones and assessing path capacity and battery levels ensure optimal performance. Once an optimal path is confirmed, data transmission occurs, followed by updating drone metrics post-transmission. However, a comprehensive surveillance system design encompasses various considerations, including drone types, operational environments, and mission-specific requirements, demanding nuanced planning beyond this outlined process. The flow chart in Fig. 2 visually represents the real-time traffic management process.

The simulation of the UAV-based surveillance using load aware multipath routing is implemented as follows:

### A. Determining the Shape of the Surveillance Area from Google Maps and Plot it

To visualize a shape using Google longitude and latitude coordinates, the geographical points can be plotted on a map. They are transformed into a normalized coordinate system for a clear representation.

### B. Image Processing

A multi-step approach is employed to perform image processing tasks, such as converting an image to grayscale, applying thresholding, and creating a binary matrix. The first step involves the conversion of the input image into grayscale. Grayscale images represent each pixel with a single intensity value, eliminating color information. Applying a filter matrix involves convolving the matrix with the grayscale. The filter matrix serves as a template that moves across the image. The covered area refers to the spatial extent over which the filter is applied.

### C. Drone Deployment

Next, determine the centroid of the entire UAV's system. This is achieved by calculating the centroid coordinates for each drone based on its spatial distribution within the shape. Once the drone centroids are identified, distances between them are computed, forming a distance matrix.

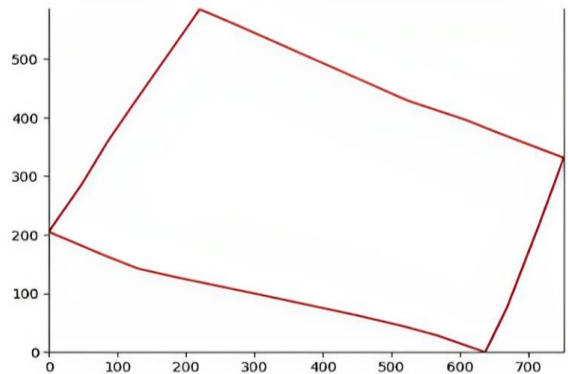

Fig. 3. Plotting reference area in the required dimension

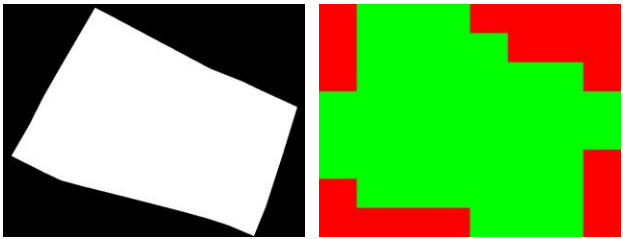

Fig. 4. Determining the coverage area of drone

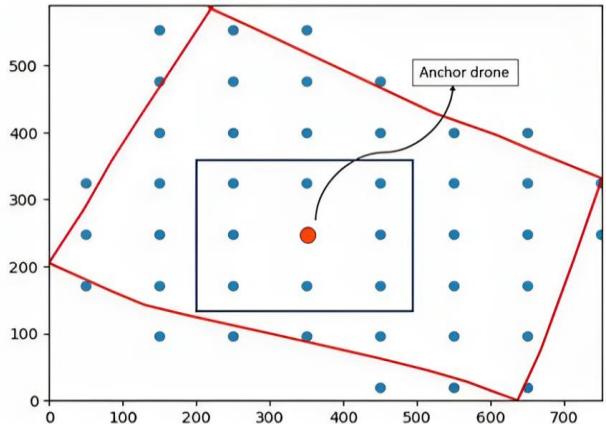

Fig. 5. Determining the position of drone to be placed and anchor drone position

### D. Path Planning

To optimize communication in a drone network, the first step involves computing pathloss and signal-to-noise ratios (SNR) for each drone pair. Pathloss quantifies the attenuation of the signal as it travels through the environment, while SNR provides a measure of the signal quality relative to background noise. Using these metrics, a drone graph is constructed, wherein nodes represent drones and edges signify communication links. The SNR values on the edges determine the strength of potential connections.

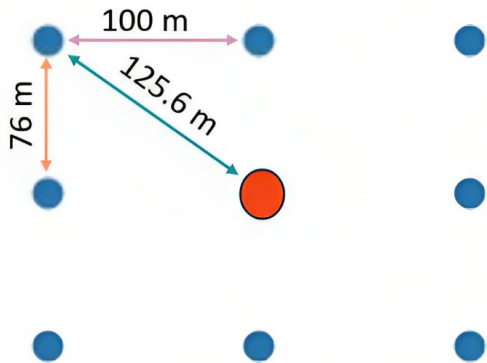

Fig. 6. Calculating distance between adjacent nodes

### V. SIMULATION RESULT

The outcomes showed that ELAMR changed into capable of correctly balance the network load via distributing the records transmission responsibilities many of the drones based on their channel capacity and battery degree. This ended in a significant discount in quit-to-quit put off and an improvement in the packet transport ratio, thereby enhancing the general performance and satisfactory of the surveillance provider. Furthermore, the multipath construction algorithm hired via ELAMR changed into successful in figuring out the most excellent paths for data transmission. By selecting paths with minimum load and maximum hyperlink quality, the algorithm ensured that the surveillance facts become transmitted in the most green way possible.

As per the image shown above, when 10% of the UAV are activated, 4 UAV gets activated. The 4 UAV activated are UAV no. 4, 12, 29 and 40. The possible combination of these UAV are determined using the above-mentioned methodology. The path determination is done with the help of channel capacity of each data transmission link.

We want to transmit the 600MB data to the path 4-8-15-23-24 and the initial capacity of the path was 3.5-3.5-3.4-3.5 Mbps respectively and initial battery level was 100% of each drone. So, to transmit 600 MB data it takes 172-172-176172 seconds. After completion of this transmission drone battery level goes to 31%-31%-32%-31%. Now, the current channel capacity is 1.085-1.085-1.088-1.085 Mbps. The same methodology is followed for another drone. The mentioned above depicts the factors undertaken and observed during transmission of the surveillance data. The factors to observed are size of the data transmitted, Path through which it gets transmitted, Channel capacity and battery life before data transmission, transmission time taken, channel capacity and remaining battery life after transmission.

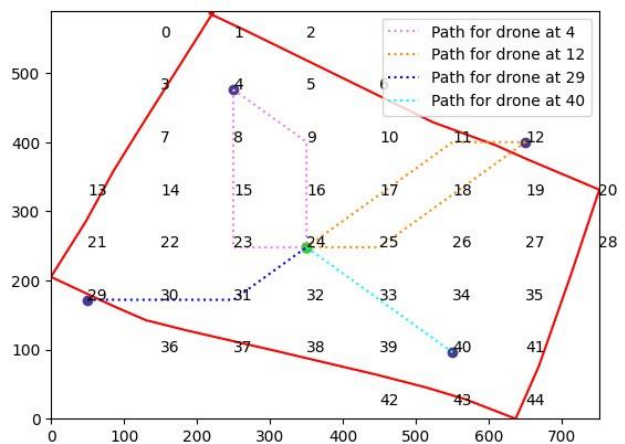

Fig. 7. Optimal path for the activated UAVs

These factors help to improve the network functionality and manage the network. The factors can be calculated with the help of above-mentioned methodology. These factors also assist when the network will be drained and how much time is required of each size of transmission data.

Table 1. Observation results before and after transmission

| Data To be Transmitted (MB) | Path | Previous capacity (Mbps) | Transmission Time (s) | Initial Battery Level | Remaining Battery level (%) | Current Capacity |
|---|---|---|---|---|---|---|
| 600 | 4->8->15->23->24 | 3.5->3.5->3.4->3.5 | 172->172->176->172 | 100->100->100->100 | 31->31->32->31 | 1.085->1.085->1.088->1.085 |
| 600 | 12->18->25->24 | 3.4->3.4->3.3 | 176->176->180 | 100->100->100 | 32->32->34 | 1.088->1.088->1.122 |
| 330 | 29->30->31->24 | 3.5->3.5->3.3 | 172->172->180 | 100->100->100 | 31->31->34 | 1.085->1.085->1.122 |
| 300 | 40->33->24 | 3.4->3.4 | 176->176 | 100->100->100 | 32->32 | 1.088->1.088 |

## VI. CONCLUSION

The approach of strategically placing drones to cover the given area with zero blind spots ensures efficient data transmission. The use of an anchor drone at the centroid helps in optimizing the network. Log Distance Path Loss and Uplink Path Loss calculations provide insights into the signal propagation characteristics. Signal-to-Noise Ratio and channel capacity calculations help in assessing the quality of communication. Activating drones in segments optimizes resource usage. Considering and choosing the most optimum path for data transmission enhances overall system efficiency. It helps to monitor battery life at each stage of data transmission. This methodology ensures reliable and efficient data transmission in a drone network, making it suitable for various applications.

## VII. FUTURE WORK

The integration of machine learning in drone technology is a significant advancement, allowing for continuous analysis of historical data, weather conditions, and user patterns to optimize drone paths. Real-time reconfiguration mechanisms are developed to respond to live traffic patterns and communication demands, ensuring efficient resource utilization. 5G technologies are explored for enhanced communication speed and reliability. The system is scalable for large coverage areas, involving multiple anchor drones and optimized communication strategies. Adaptive bandwidth allocation mechanisms are used to optimize performance under diverse conditions. Edge computing capabilities on drones are integrated to reduce latency and minimize reliance on centralized processing. Advanced algorithms for autonomous navigation and collision avoidance enhance safety measures, ensuring a robust and secure system. This comprehensive approach leverages historical data, scalability, and autonomy to create a resilient and efficient drone network.